\documentclass[twocolumn]{aastex62}

\newcommand{\feh}{$\mbox{[Fe/H]}$}

\received{February, 2018}
\revised{March, 2018}
\accepted{March, 2018}

\submitjournal{ApJ}

\shorttitle{Very metal-poor stars in the Sagittarius dSph}
\shortauthors{Chiti \& Frebel}

\begin{document}

\title{Four Metal-poor Stars in the Sagittarius Dwarf Spheroidal Galaxy\footnote{This paper includes data gathered with the 6.5\,m Magellan Telescopes located at Las Campanas Observatory, Chile.}}

\correspondingauthor{Anirudh Chiti}
\email{achiti@mit.edu}

\author[0000-0002-7155-679X]{Anirudh Chiti}
\affil{Department of Physics and Kavli Institute for Astrophysics and Space Research, Massachusetts Institute of Technology, Cambridge, MA 02139, USA}

\author[0000-0002-2139-7145]{Anna Frebel}
\affiliation{Department of Physics and Kavli Institute for Astrophysics and Space Research, Massachusetts Institute of Technology, Cambridge, MA 02139, USA}

\begin{abstract}

We present the metallicities and carbon abundances of four newly discovered metal-poor stars with $ -2.2 <\,\,\feh\,\, < -1.6$ in the Sagittarius dwarf spheroidal galaxy. 
These stars were selected as metal-poor member candidates using a combination of public photometry from the SkyMapper Southern Sky Survey and proper motion data from the second data release from the \textit{Gaia} mission.
The SkyMapper filters include a metallicity-sensitive narrow-band $v$ filter centered on the Ca II K line, which we use to identify metal-poor candidates.
In tandem, we use proper motion data to remove metal-poor stars that are not velocity members of the Sagittarius dwarf spheroidal galaxy.
We find that these two datasets allow for efficient identification of metal-poor members of the Sagittarius dwarf galaxy to follow-up with further spectroscopic study.
Two of the stars we present have $\feh\,\,< -2.0$, which adds to the few other such stars currently identified in the Sagittarius dwarf galaxy that are likely not associated with the globular cluster M54, which resides in the nucleus of the system.
Our results confirm that there exists a very metal-poor stellar population in the Sagittarius dwarf galaxy. 
We find that none of our stars can be classified as carbon-enhanced metal-poor stars.
Efficiently identifying members of this population will be helpful to further our understanding of the early chemical evolution of the system.

\end{abstract}


\section{Introduction} 
\label{sec:intro}

Studying the detailed chemical abundances of metal-poor stars\footnote{Defined as $\feh\,\,\le -1\,\text{dex}$, where [Fe/H] = $\log_{10}(N_{\text{Fe}}/N_{\text{H}})_{\star}-\log_{10}(N_{\text{Fe}}/N_{\text{H}})_{\sun}$ \citep{bc+05, fn+15}.} in our own galaxy, the Milky Way, allows us to probe the physical processes that governed element formation in the early universe.
For instance, the most metal-poor stars can be used to study the yields of early nucleosynthesis events \citep[i.e.,][]{un+03, ntu+06, hw+10, pfb+16}.
Studying these enrichment events helps constrain the properties (e.g., initial mass function) of the First Stars \citep[e.g.,][]{itk+18}.
Additionally, comparing the detailed chemical abundance patterns of metal-poor stars in the Milky Way halo to those in other environments such as dwarf galaxies can help constrain paradigms of galaxy formation and environment-related dependencies in star formation and chemical enrichment \citep[reviewed in][]{ths+09, fn+15}. 
The efficient identification of metal-poor stars in a variety of environments is a necessary prerequisite for the aforementioned studies. 

Early work on the detection of metal-poor halo stars relied on measuring the strength of the Ca II K absorption line at 3933.7\,\AA\,\,in large samples of low-resolution and medium-resolution stellar spectra \citep{bps+85, bps+92}.
Then, more detailed spectroscopic follow-up of the most promising candidates would be performed.
This technique of selecting metal-poor candidates from samples of low-resolution or medium-resolution spectra, and then conducting follow-up observations, has been replicated in a number of large surveys such as the Hamburg-ESO Survey \citep{c+03, fcn+06}, the Sloan Extension for Galactic Understanding and Exploration \citep[SEGUE;][]{lbs+08, asb+08, abl+13, aag+16}, the Radial Velocity Experiment \citep[RAVE;][]{fwr+10}, and survey work with the LAMOST telescope \citep{aga+17, ltz+18}.
The success of these surveys has led to the discovery of more than 500 extremely metal-poor (EMP) stars, which are defined as having [Fe/H] $< -3.0$ \citep[][and references therein]{af+18}.

Recently, narrow-band photometry has been used to identify metal-poor candidates \citep[i.e.,][]{ksb+07, smy+17, wpb+18}.
This involves using a narrow-band $v$-filter encompassing the region of the Ca II K line as the flux through the narrow-band filter is strongly related to the strength of the Ca II K line.
Hence the flux can be related to the overall metallicity of the star, in particular for metal-poor stars.
The gain from this technique over spectroscopic identification is that photometry requires less observing time than spectroscopy, and the ability to simultaneously derive metallicity information on all stars to a given magnitude.
So far, the application of this technique in the SkyMapper Southern Sky Survey has led to the discovery of over 100 EMP stars \citep[e.g.,][]{jkf+15} and a star with an upper limit on the iron abundance of [Fe/H] $< -6.5$ \citep{kbf+14, nal+17}.
Recent work by the Pristine Survey with a more finely tuned narrow-band $v$-filter has led to the discovery of a star with [Fe/H] $ = -4.7$ \citep{sab+18}, among others.

A natural venue in which to apply this selection techniques are dwarf spheroidal galaxies (dSphs). 
Only in the past decade have stars with [Fe/H] $< -3.0$ been discovered in dwarf galaxies \citep{kgs+10, fks+10} which are simpler systems than the Milky Way due to their smaller size and limited star formation history.
Their localized context facilitates interpretation when relating the chemical abundances of metal-poor stars to properties of the galaxy (i.e.,  star formation history, chemical enrichment events).
Certainly, the chemical abundances of EMP stars in dSphs appear to show many similarities to the Milky Way halo population, in accordance with current paradigms of hierarchical galaxy formation (see \citealt{fn+15} for a review).
Further work on efficiently identifying the most metal-poor stars in any dwarf galaxy would thus be useful in probing these similarities across as many systems as possible.

In this paper, we choose to use a combination of SkyMapper \citep{wol+18} and \textit{Gaia} public data \citep{gaia+16, gaia+18} to implement an efficient technique to identify metal-poor stars in the Sagittarius dSph \citep{igi+94}. 
Early work on the chemical abundances of stars in the Sagittarius dSph focused primarily on stars with [Fe/H] $> -1.6$ \citep{bhm+00, bsm+04, mbb+05, sbb+07, bic+08, cbg+10, mwm+13, hss+17}.
\citet{bic+08} found four stars with [Fe/H] $< -2.0$ in the nucleus of the Sagittarius dSph, one of which was recovered in the sample of \citet{mbi+17}.
However, the proximity of these stars to the globular cluster M54, which also lies in the nucleus of the Sagittarius dSph, make their association with the main body of the Sagittarius dSph slightly less clear.
Recently, \citet{hem+18} published detailed chemical abundances for three more stars with [Fe/H] $< -2.0$ in the Sagittarius dSph that are beyond the tidal radius of M54 and are thus associated with the main body of the Sagittarius dSph.
The chemical abundances of these stars show some similarities to stars with [Fe/H] $< -2.0$ in the Milky Way halo, and have disputed the nature of the Sagittarius dSph having a top-light initial mass function as argued from the more metal-rich population \citep[e.g.,][]{hss+17}.
By adding to the population of the most metal-poor stars known in the Sagittarius dwarf galaxy, we aim to probe its early chemical and assembly history. 
Here we report the discovery of four photometrically-selected metal-poor stars in the Sagittarius dSph.
We present spectroscopic measurements of the [Fe/H] and carbon abundance of these stars, of which two have $\feh\,\,< -2.0$.

The paper is organized as follows. 
In Section~\ref{sec:obs}, we outline our target selection procedure and observations; in Section~\ref{sec:analysis}, we present our analysis in deriving the chemical abundances of these stars; in Section~\ref{sec:results}, we discuss the chemical abundance signatures of these stars, the efficiency of our target selection procedure, and overall findings regarding the early history of the Sagittarius dSph; in Section~\ref{sec:conclusion}, we provide summarize our results.

\begin{deluxetable*}{lllrrrrr}[!htbp] 
\tablecaption{\label{tab:obs} Observations}
\tablehead{   
  \colhead{Name} &
  \colhead{RA (h:m:s) (J2000)} & 
  \colhead{DEC (d:m:s) (J2000)} &
  \colhead{Slit size} &
  \colhead{$g$ (mag)} &
  \colhead{$t_{\text{exp}}$ (min)} &
  \colhead{S/N$\tablenotemark{a}$} & 
  \colhead{$v_{\text{helio}}$ (km/s)}}
\startdata
Sgr-2 & 19:01:39.16 & $-32$:56:44.1 & 1\farcs0 & 16.83 & 15 & 22, 35 & 142.7\\ 
Sgr-7 & 18:50:32.63 & $-32$:35:34.0 & 1\farcs0 & 16.81 & 20 & 30, 60 & 169.8\\ 
Sgr-9 & 18:55:51.59 & $-30$:39:45.4 & 1\farcs0 & 17.10 & 15 & 25, 45& 142.6\\ 
Sgr-10 & 18:50:23.40 & $-31$:09:00.1 & 1\farcs0 & 15.96 & 6 & 20, 40& 173.0 
\enddata
\tablenotetext{a}{S/N per pixel is listed for 4500\,\AA\,\,and 8500\,\AA}
\end{deluxetable*}

\section{Target Selection \& Observations}
\label{sec:obs}

Traditionally, candidate stellar members of a dwarf galaxy are identified in a color-magnitude diagram along an isochrone.
Then, spectra are obtained for these stars to determine their membership status based on metallicity and velocity measurements.
With our new technique, we increased the efficiency of the target-selection procedure by leveraging publicly available data to select metal-poor star candidates that appear to have similar proper motions and thus might be associated with a common dwarf galaxy system.

\begin{figure*}[!htbp]
\centering
\includegraphics[width =1.00\textwidth]{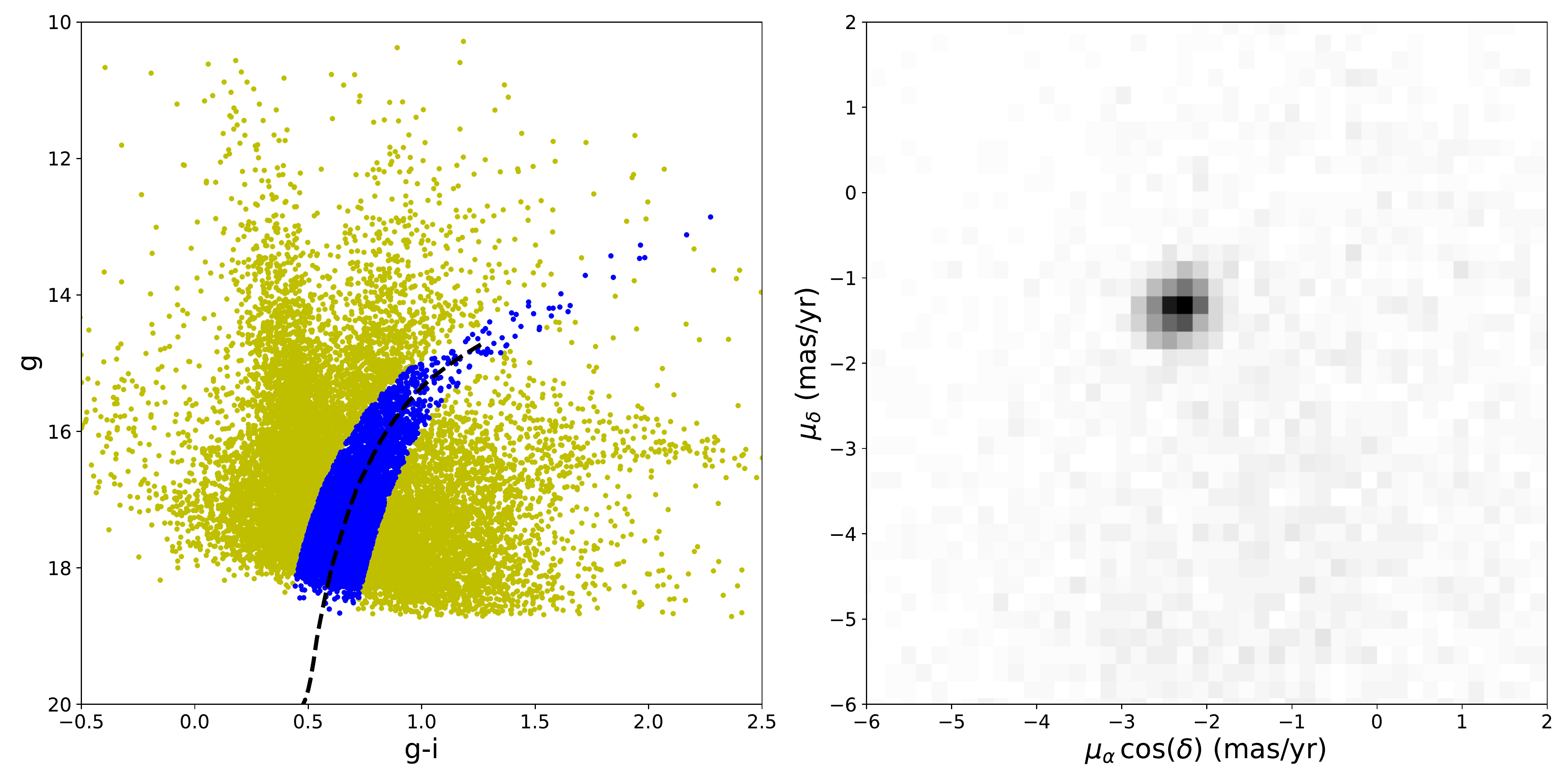}
\caption{Left: Color-magnitude diagram of all sources within 60\arcmin of the center of the Sgr dSph.
A 10 Gyr, [Fe/H] = $-2.0$ Dartmouth isochrone is overlaid \citep{dcj+08}, and points within $g-i \pm 0.15$ are marked in blue. 
Right: \textit{Gaia} DR2 proper motions of the blue data points in the left panel. 
An over-density in proper motion space is centered on $\mu_\alpha\cos(\delta)=-2.33\,\text{mas/yr}, \mu_\delta=-1.36\,\text{mas/yr}$, indicating the systemic proper motion of the Sgr dSph members.}
\label{fig:CMD}
\end{figure*}

\subsection{Target Selection}
\label{sec:targets}

We queried the SkyMapper DR1.1 catalog \citep{wol+18} to retrieve photometric information on all sources within 2.85$^{\circ}$ of the center of the Sagittarius dSph ($\alpha_{\text{J2000}}$ = 18h55m19.5s, $\delta_{\text{J2000}} =-30$d32m43s).
We opted to use petrosian magnitudes, denoted in the source catalog by the {\it \_petro} flag, for all subsequent analysis.
The photometry was de-reddened following the prescription in \citet{wol+18} using reddening maps from \citet{sfd+98}.
To remove sources that are likely not stars, we chose to exclude all sources with the catalog flag {\it class\_star} $<$ 0.9.

From the resulting catalog, we used $g-i$ colors and $g$ magnitudes to construct a color-magnitude diagram (CMD) to use for a first-pass selection of candidate members of the Sagittarius dSph.
A 10\,Gyr, [Fe/H] = $-2.0$ isochrone from the Dartmouth Stellar Evolution Database \citep{dcj+08} was overlaid on the CMD at the distance modulus of the Sgr dSph \citep[16.97;][]{kc+09}.
All stars within $(g-i)\pm 0.15$ of the isochrone were kept as candidate members.
The result of this procedure is shown in the leftmost panel of Figure~\ref{fig:CMD}.

We then retrieved proper motion data from the {\it Gaia} DR2 catalog \citep{gaia+16, gaia+18, sgg+17} for all stars within $(g-i)\pm 0.15$ of the isochrone.
Upon plotting each star in proper motion space, we found an over-density centered on $\mu_\alpha\cos(\delta)=-2.33\,\text{mas/yr}, \mu_\delta=-1.36\,\text{mas/yr}$ as shown in the right panel in Figure~\ref{fig:CMD}.
Since this over-density is clearly distinct from the foreground population and since stellar members of a dSph should have similar velocities, it therefore likely corresponds to stellar members of the Sagittarius dSph.
Furthermore, the location of this over-density agrees well with the proper motion measurement of the Sagittarius dSph derived using \textit{Gaia} DR2 data in \citet{ghv+18}.
We then conservatively narrowed our selection of candidate members to all stars with proper motion measurements $-2.75\,\text{mas/yr} < \mu_\alpha\cos(\delta) < -1.90\,\text{mas/yr}$, and $-1.75\,\text{mas/yr} < \mu_\delta < -1.0\,\text{mas/yr}$.
 
To select the most metal-poor of these stars, we leveraged the $v$-band photometry in the SkyMapper DR1.1 catalog.
The SkyMapper narrow-band $v$-filter is sensitive to stellar metallicities because much of the bandpass of the filter is encompassed by the prominent Ca II K line \citep{bbs+11}. 
We plotted our candidate members in $v-g-0.9\times(g-i)$ vs. $g-i$ space, since metal-poor stars tend to have lower $v-g-0.9\times(g-i)$ values for a given $g-i$.
This argument is based on work presented in \citet{kss+12}, who used an index of $v-g-2\times(g-i)$.
Since this differs from our index, we had to verify whether the most metal-poor stars will also lie at lower $v-g-0.9\times(g-i)$ indices.
We therefore applied this selection process to several globular clusters in the footprint of the SkyMapper survey to test whether their member stars were identifiable as more metal-poor than foreground stars.

\begin{figure*}[!htbp]
\centering
\includegraphics[width =0.70\textwidth]{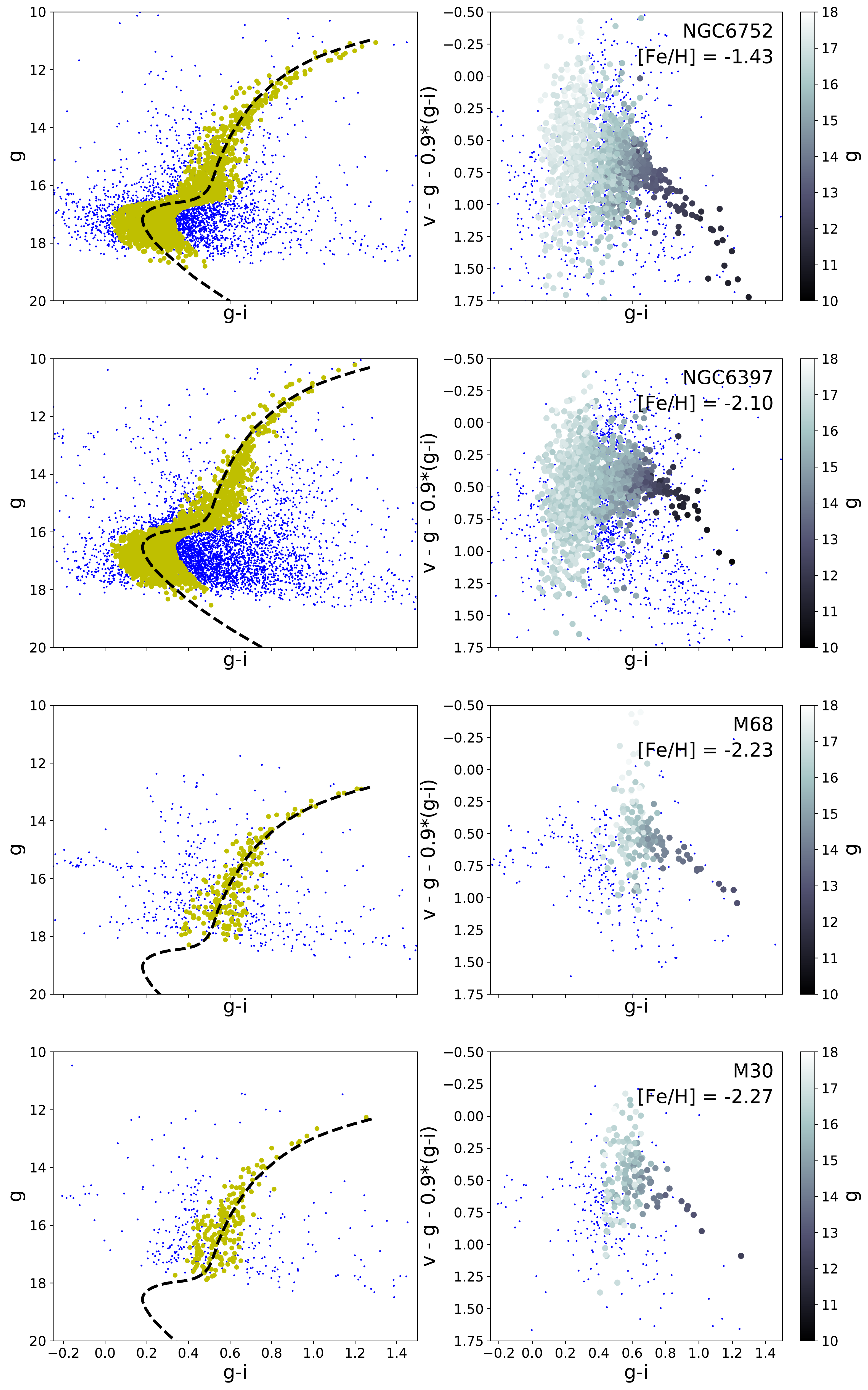}
\caption{Left: Color-magnitude diagrams of NGC6752, NGC6397, M68, and M30, from top to bottom of stars within 15$\arcmin$ from their centers. 
Dartmouth isochrones of 10 Gyr, [Fe/H] = $-2.0$ are overlaid. 
Right: Metallicity-sensitive color-color plots using SkyMapper photometry of the corresponding globular clusters, with the magnitudes of the data points along the isochrone color-coded by magnitude. 
As noted in Section~\ref{sec:targets}, the measurements appear to start being less sensitive to metallicity at a magnitude $g \sim 16$, due to a lack of photometric precision.}
\label{fig:metallicities}
\end{figure*}

We performed the same CMD and proper motion selection procedure on four globular clusters: NGC6752
\citep[\text{[Fe/H] =} $-1.43$;][]{smd+17}, NGC6397 \citep[\text{[Fe/H] =} $-2.10$;][]{kw+11}, M68 \citep[\text{[Fe/H] =} $-2.23$;][]{cbg+09}, M30 \citep[\text{[Fe/H] =} $-2.27$;][]{cbg+09}.
Since all member stars of a globular cluster have a similar metallicity, their members should form a distinct contour in $v-g-0.9\times(g-i)$ vs. $g-i$ space.
In Figure~\ref{fig:metallicities}, we see a clear separation in $v-g-0.9\times(g-i)$ vs. $g-i$ space between foreground halo stars and members of the globular clusters NGC6397, M68, and M30.
The separation is visible, but less prominent, for the member stars of NGC6752, since they do not separate as clearly from the foreground.
This result implies that stars with metallicities below [Fe/H] $=-1.43$, which is the metallicity of NGC6752, should begin to separate from the halo foreground in our metallicity selection.
Of note, we also found that SkyMapper DR.1.1 $v$ band photometry for stars fainter than $g\sim16$ appears to have insufficient precision to clearly separate members from the foreground.
Thus, having demonstrated the utility of selecting metal-poor candidates using this technique, we followed this procedure to pick metal-poor candidate stars in the Sagittarius dSph. 
Following \citet{hem+18}, we observed stars outside the tidal radius of the nearby globular cluster, M54 \citep[7\farcm5;][]{tkd+95} to ensure our targets were not members of that system.

\subsection{Observations \& Data Reduction}
\label{sec:data}

We used the Magellan Echellette (MagE) Spectrograph \citep{mbt+08} on the Magellan-Baade telescope at Las Campanas Observatory to obtain medium-resolution spectra of six metal-poor candidate member stars in the Sagittarius dSph.
These stars were selected since their $v-g-0.9\times(g-i)$ indices were among the lowest in the sample.
Targets were observed with the 1\farcs0 slit and 1x1 binning, which grants sufficient resolution ($R\sim4000$) and wavelength coverage (3200\,\AA\,\,$-$ 10000\,\AA) to derive abundances from the Ca II K line ($\sim3933$\,\AA), the CH G band ($\sim4300$\,\AA), the Mg b region ($\sim5150$\,\AA), and the calcium triplet lines ($\sim8500$\,\AA). 
We observed these stars for the first $\sim$1.5 hours of the night of July 17th, 2018, during which the weather was partially cloudy.

Our spectra were reduced using the Carnegie Python pipeline \citep{k+03}\footnote{https://code.obs.carnegiescience.edu/mage-pipeline} using standard calibration procedures.
To account for potential effects from instrument stability on the wavelength calibration, a ThAr calibration arc lamp spectrum was collected after slewing to each target.
Spectra of each target were then reduced using the corresponding arc lamp spectrum. 
Four of the six observed stars were later determined to be members of the Sagittarius dSph.
Details of their observations are shown in Table~\ref{tab:obs}.

\section{Analysis}
\label{sec:analysis}

Here we present our methods for deriving the stellar parameters ($T_{\text{eff}}$, $\log g$), metallicities, and carbon abundances of the stars we observed.
The $T_{\text{eff}}$ and $\log g$ values were derived by matching public photometry from SkyMapper DR1.1 to isochrones from the Dartmouth Stellar Evolution Database \citep{dcj+08}.
The metallicity was derived based on measurements of the strengths of the Ca II K line ($\sim3933.7$\,\AA), Mg b region ($\sim5150$\,\AA), and calcium triplet lines ($\sim8500$\,\AA).
The carbon abundance was derived from the CH G band ($\sim4300$\,\AA).

In subsequent analysis, we assume the stars in our sample are members of the red giant branch.
There exists a chance that some are instead asymptotic giant branch (AGB) stars, due to the proximity of the AGB track to the red giant branch.
However, the timescale of stars existing on the AGB track is short compared to the timescale that stars exist on the red giant branch, which makes it unlikely that any stars in our sample of only four stars are AGB stars.

\subsection{Radial Velocity Measurements \& Deriving Stellar Parameters}
\label{sec:stellarparams}
As a preliminary step, it was necessary to measure the radial velocity of each star to verify it as a member of the Sagittarius dSph.
To derive radial velocities, we cross-correlated each spectrum with a template spectrum of the metal-poor giant HD122563 in the Spectroscopy Made Hard (SMH) analysis software \citep{c+14}.
Heliocentric velocity corrections were derived using the task \texttt{rvcorrect} in \texttt{IRAF}.
We found that the four stars in Table~\ref{tab:obs} have velocities consistent with the systemic velocity of the bulge of the Sagittarius dSph of $\sim141$\,km\/s $\pm$ 9.6\,km\/s \citep{bic+08}.
While this comparison may not be strictly applicable to Sgr-2, Sgr-7, and Sgr-10 since they are over one degree from the center of the Sagittarius dSph, we note that their velocities are comparable to the velocities of the population of metal-poor members of the Sagittarius dSph presented in \citet{hem+18}, which range from 127.4\,km/s to 167.2\,km/s.
The other two stars that we observed had radial velocities $\sim180$\,km/s below the systemic velocity of the bulge of the Sagittarius dSph, and we thus concluded that they were not members of the system.

The stellar parameters, $T_{\text{eff}}$ and $\log g$, were derived by matching SkyMapper $g$ and $i$ photometry to a 10 Gyr, [Fe/H] = $-2.0$ isochrone from the Dartmouth Stellar Evolution Database.
To validate this method of measuring stellar parameters, we performed the same procedure on the seven metal-poor and very metal-poor Sagittarius dSph stars with stellar parameter measurements in \citet{hem+18} that also had publicly available SkyMapper photometry. 
\citet{hem+18} derived $T_{\text{eff}}$ and $\log g$ from high-resolution spectra by removing trends in excitation potential with abundance and satisfying ionization equilibrium.
For these seven stars, we derive $T_{\text{eff}}$ values that are consistent with the measurements in \citet{hem+18}.
We measure a marginally higher $T_{\text{eff}}$ of 24\,K on average and the residuals have a standard deviation of 176\,K.
For the $\log g$ values, our measurements are larger than those in \citet{hem+18} by 0.3\,dex on average and the residuals have a standard deviation of 0.7\,dex.

To further test the validity of our stellar parameter measurements, we compared our $T_{\text{eff}}$ values to those predicted from a color-$T_{\text{eff}}$-[Fe/H] relation.
We converted the SkyMapper photometry to the SDSS photometric system by first converting to the Pan-STARRS photometric system (described in paragraph 2 of Section~\ref{sec:ca2k}) and then the SDSS photometric system using the color transformations in \citet{tsl+12}.
Then, we applied the IRFM temperature estimator\footnote{https://www.sdss.org/dr14/spectro/sspp\_irfm/}, which is an additional photometric $T_{\text{eff}}$ calibration that was added to the original SEGUE Stellar Parameter Pipeline \citep{ybs+08}.
The IRFM temperature estimator gives largely reasonable $T_{\text{eff}}$ values that are offset from our original $T_{\text{eff}}$ values by $-300$\,K (Sgr-2), 40\,K (Sgr-7), and $-71$\,K (Sgr-10).
For Sgr-9, the estimator gives a $T_{\text{eff}}$ value that is 590\,K lower than our original measurement.
We choose to adopt our original $T_{\text{eff}}$ value of 4767\,K for Sgr-9, since a visual comparison of the Balmer lines in its spectrum to the Balmer lines in the spectra of the other Sagittarius dSph members suggests that all have similar $T_{\text{eff}}$ values.

We also compared our $\log g$ values to those derived from an estimate of the absolute magnitude of each of our stars by assuming each star has a distance modulus equal to that of the Sagittarius dSph \citep[16.97;][]{kc+09}.
We then derive $\log g$ values using the canonical equation presented in \citet{hem+18}, assuming a mass of 0.7\(\textup{M}_\odot\) and the $T_{\text{eff}}$ values in Table~\ref{tab:measurements}.
We find our $\log g$ values are 0.4\,dex on average below those derived by the above method, with a standard deviation of the differences in $\log g$ values between the two methods of 0.15\,dex.

\subsection{Deriving Chemical Abundances}

We derived the metallicity ([Fe/H]) of each star in our sample from the equivalent widths of the Ca II K line (3933.7\,\AA) and the calcium triplet lines (8498\,\AA, 8542\,\AA, and 8662\,\AA).
We used these measurements to derive metallicities by applying calibrations detailed in \citet{brn+99} and \citet{cpg+13}, respectively.
Their application is further described in Sections~\ref{sec:ca2k} and~\ref{sec:caT}.
By using spectral synthesis techniques, we also independently derived the metallicity using the Mg b region ($\sim5150$\,\AA).
We derived a carbon abundance ([C/Fe]) for each star using the CH G band ($\sim4300$\,\AA).
Details of methodology are provided in Sections~\ref{sec:mgb} and \ref{sec:gband}.

\begin{figure*}[!htbp]
\centering
\includegraphics[width =\textwidth]{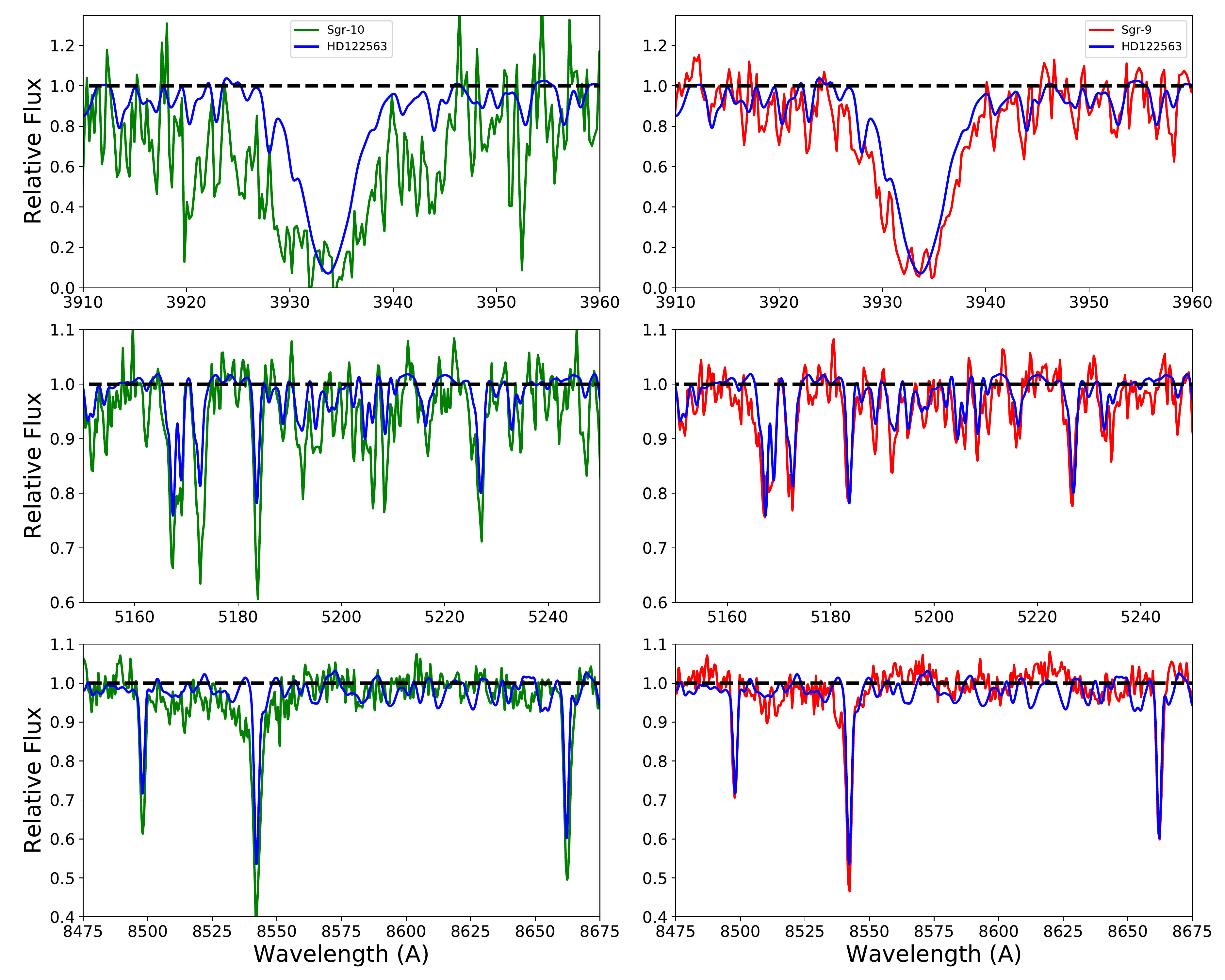}
\caption{Sample spectra of the Ca II K line (top), Mg b region (middle), and the calcium triplet region (bottom). 
Spectra for Sgr-10 ([Fe/H] = $-1.70$) are shown in green on the left, and those for Sgr-9 ([Fe/H] = $-2.3$) on the right.
In all plots, a MIKE spectrum of HD122563 \citep[\text{[Fe/H] = }$-2.64$;][]{jhs+14} smoothed to $R\sim4000$ is overplotted.}
\label{fig:specs}
\end{figure*}

We took the weighted average of the three iron abundances from the methods described in Sections~\ref{sec:ca2k},~\ref{sec:caT}, and~\ref{sec:mgb} to derive final [Fe/H] values.
Examples of our spectra covering the three wavelength regions used to derive [Fe/H] are presented in Figure~\ref{fig:specs}.
All chemical abundance measurements are presented in Table~\ref{tab:measurements} and are listed relative to solar abundances from \citet{ags+09}.

\subsubsection{Ca II K Line}
\label{sec:ca2k}

\citet{brn+99} presented a calibration that related a star's $B-V$ color and the strength its Ca II K line at 3933.7\,\AA\,\,to metallicity.
The calibration used the KP index, a measurement of the pseudo-equivalent width of the Ca II K line, to quantify line strength.
A full discussion of deriving the KP index, which involves measuring separate line indices known as K6, K12, and K18, is presented in \citet{brn+99}. 
We exactly followed their implementation.
For continuum normalization, we fit a line through the blue and red sidebands of the Ca II K line located between 3903\,\AA$-$3923\,\AA\,\,and 4000\,\AA$-$4020\,\AA, respectively.
We then directly integrated over the bandpasses presented in \citet{brn+99} to measure the K6, K12, and K18 line indices, and adopted a final KP index following their prescription.
The uncertainty on the KP index was derived by shifting the continuum placement based on the signal-to-noise of the spectrum.

To obtain $B-V$ colors for each of our stars, we first derived a conversion between SkyMapper $g$ and $r$ photometry and the Pan-STARRS photometric system using data for K, G, and F type stars on the SkyMapper Southern Sky Survey website\footnote{http://skymapper.anu.edu.au/filter-transformations} based on the spectral library from \citet{p+98}.
We then converted the Pan-STARRS $g$ and $r$ photometry to $B-V$ colors using color transformations provided in \citet{tsl+12}.
The resulting uncertainties on the $B-V$ colors from these transformations ranged from 0.04\,mag and 0.08\,mag.

Additionally, we note that \citet{brn+99} assumes [Ca/Fe] = 0.4 when [Fe/H] $< -1.5$, which holds true for stars in the Milky Way halo.
Results from \citet{mbi+17} and \citet{hem+18} suggest general agreement between the [Ca/Fe] values of Milky Way halo stars and Sagittarius dSph stars for metal-poor stars.
Thus, we apply no further correction to our \feh\,\,values derived from the calibration.

The uncertainty on the [Fe/H] from this method was derived by adding in quadrature the model uncertainties from the \citet{brn+99} calibration and the shifts in [Fe/H] by varying the KP index and $B-V$ colors by their uncertainties.

\subsubsection{Ca Triplet Lines}
\label{sec:caT}
The equivalent widths of the calcium triplet lines at 8498\,\AA, 8542\,\AA, and 8662\,\AA\,\,can also be related to the overall metallicity of the star, as detailed in \citet{cpg+13}.
We measured the equivalent width of each of these lines in our spectra using the \texttt{splot} function in \texttt{iraf}. The uncertainty on each of these equivalent width measurements was determined by varying the continuum placement.
We note that for Sgr-2, fringing near the calcium triplet lines in its spectrum led to higher uncertainties.

We opted to use the form of the calibration in \citet{cpg+13} that requires an absolute $V$ magnitude measurement.
To derive this value, we followed the same steps outlined in the second paragraph of Section~\ref{sec:ca2k} to convert SkyMapper photometry to the Pan-STARRS photometric system.
We then applied transformations from \citet{tsl+12} to derive $V$ magnitudes. 
Then, we derived an absolute $V$ magnitude by subtracting the distance modulus of the Sagittarius dSph \citep[16.97;][]{kc+09}.

The uncertainty on the [Fe/H] was derived by propagating the model uncertainties in \citet{cpg+13} calibration, and adding them in quadrature to the shifts in [Fe/H] by varying the equivalent widths and $V$ by their respective uncertainties.

\subsubsection{Mg b Region}
\label{sec:mgb}
We fit synthetic spectra to the Mg b region (5150\,\AA $-$ 5220\,\AA) to derive metallicities.
All syntheses and fitting were performed with the SMH software using the 2017 version of the 1D LTE radiative transfer code MOOG \citep{s+73}\footnote{https://www.as.utexas.edu/~chris/moog.html} with an updated treatment of scattering \citep{sks+11}\footnote{https://github.com/alexji/moog17scat} and the Kurucz model atmospheres \citep{cf+04}.
The line list was compiled using software provided by C. Sneden.
The software retrieved data from the \citet{k+11} database and added measurements from references in \citet{slc+09, slr+14, sck+16}. 
The $T_{\text{eff}}$ and $\log g$ values derived in Section~\ref{sec:stellarparams} were used as stellar parameters when synthesizing spectra.
A microturbulence of v$_{\text{micro}} = 2.0$ was assumed in all syntheses.

We generated synthetic spectra with different [Fe/H] until there was agreement with the observed spectra.
The [Mg/Fe] ratio was fixed to 0.4, following from a general agreement in [$\alpha$/Fe] values between Milky Way halo stars and metal-poor Sagittarius dSph stars \citep{mbi+17, hem+18}.

To derive the uncertainty on our values, we added random and systematic sources of uncertainty in quadrature.
The random uncertainty was derived by noting the variation in [Fe/H] required to encapsulate the noise in each spectrum.
This procedure led to random uncertainties between 0.20\,dex and 0.25\,dex.
The systematic sources of uncertainty were from uncertainties in the stellar parameter measurements.
We find that uncertainties of $\sim$150\,K in $T_{\text{eff}}$ and $\sim$0.3 in $\log g$ arise when propagating the SkyMapper photometric uncertainties through our method of deriving stellar parameters.
We thus vary each stellar parameter by its uncertainty, re-derive [Fe/H], and note the discrepancy from the original [Fe/H] as a systematic uncertainty.
We then add all systematic uncertainties and the random uncertainty in quadrature to derive a final uncertainty.
Final derived values and uncertainties are presented in Table~\ref{tab:measurements}.

\subsubsection{G Band}
\label{sec:gband}
We fit synthetic spectra to the CH G band between 4260\,\AA\,\,and 4320\,\AA\,\,to derive carbon abundances.
The same procedure was followed as in Section~\ref{sec:mgb} with the exception of the following details.
First, the line list was compiled with the same sources as in Section~\ref{sec:mgb}, but CH data were added from \citet{mpv+14}.
Second, the [Fe/H] value in the synthetic spectra was fixed to the final [Fe/H] presented in Table~\ref{tab:measurements}, and the [C/Fe] value was allowed to vary.
Third, when calculating the uncertainties, we include the uncertainty in [Fe/H] from Table~\ref{tab:measurements} as a source of systematic uncertainty in [C/Fe]. 
Finally, the derived carbon abundances were corrected for the evolutionary state of the star following \citet{pfb+14}.
The final carbon abundance values are presented in Table~\ref{tab:measurements}.

\begin{deluxetable*}{lrrrrrrrr}
\tablecaption{\label{tab:measurements} Stellar parameters and chemical abundances}
\tablehead{
\colhead{Name} & 
  \colhead{$T_{\text{eff}}$} & 
  \colhead{log\,$g$} &
\colhead{[Fe/H]$_{\text{Mg}}$} & 
\colhead{[Fe/H]$_{\text{CaT}}$} & 
\colhead{[Fe/H]$_{\text{Ca2k}}$} & 
\colhead{[Fe/H]$_{\text{final}}$} & 
\colhead{[C/Fe]} & 
\colhead{[C/Fe]$_{\text{corrected}}$ $\tablenotemark{a}$}\\
  \colhead{} &
  \colhead{(K)} & 
  \colhead{(dex)} &
  \colhead{(dex)} & 
  \colhead{(dex)} & 
  \colhead{(dex)} & 
  \colhead{(dex)} & 
  \colhead{(dex)} & 
  \colhead{(dex)}}
\startdata
Sgr-2 & 4836 & 1.71 & $-2.20\pm0.33$ & $-1.82\pm0.43$ & $-2.25\pm0.41$ & $-2.11\pm0.18$ & $0.02\pm0.36$ & $0.34\pm0.36$ \\
Sgr-7 & 4610 & 1.25 & $-1.95\pm0.34$ & $-1.69\pm0.30$ & $-1.49\pm0.46$ & $-1.74\pm0.17$ & $-0.52\pm0.33$ & $0.04\pm0.33$ \\
Sgr-9 & 4767 & 1.56 & $-2.47\pm0.36$ & $-2.10\pm0.17$ & $-2.34\pm0.39$ & $-2.19\pm0.15$ & $-0.66\pm0.44$ & $-0.25\pm0.44$ \\
Sgr-10 & 4610 & 1.25 & $-1.84\pm0.32$ & $-1.57\pm0.27$ & $-1.54\pm0.48$ & $-1.66\pm0.13$ & $-0.37\pm0.35$ & $0.17\pm0.35$
\enddata
\tablenotetext{a}{Corrected for the evolutionary state of the star following \citet{pfb+14}.}
\end{deluxetable*}

\section{Results}
\label{sec:results}

In Section~\ref{sec:carbon}, we discuss the relatively low carbon abundances of the stars in our sample, and in Section~\ref{sec:targetsuccess}, we discuss in more detail the implementation of our technique for identifying metal-poor stars in the Sagittarius dSph.
Since our sample of four is small and biased toward metal-poor stars, further interpretation is difficult in the context of the overall metallicity distribution of stars in the system. 
However, we confirm that the metallicity distribution function reaches below $\feh\,\,= -2.0$.
Our fairly efficient detection of two very metal-poor stars builds on the previous detection of such stars in the system \citep{bic+08, mbi+17, hem+18}, and bolsters the argument that Sagittarius dSph should host a population of very metal-poor stars.
Our results suggest that this population can be identified for further study by using efficient target selection techniques, thus facilitating a more complete understanding of the early chemical evolution of the system.

\subsection{The Signature of Carbon in the Sagittarius dSph}
\label{sec:carbon}

Carbon is an important element for studying early chemical evolution, partly because the fraction of stars that are enhanced in carbon increases as metallicity decreases \citep{cbb+12, pfb+14}.
This observation has led to the identification of carbon-enhanced metal-poor (CEMP) stars ([C/Fe] $>$ 0.7, [Fe/H] $< -1.0$) as a separate subclass of metal-poor stars \citep{bc+05, abc+07}.
\citet{pfb+14} measured the fraction of CEMP stars to be $\sim20\%$ for stars with [Fe/H] $< -2.0$.
However, CEMP stars are not found in all environments.
For instance, only a few s-process rich CEMP stars have been detected in the direction of the Milky Way bulge \citep{hca+15, kmp+16} and studies generally tend to find a lower fraction of CEMP stars in dSphs \citep{kgz+15, jnm+15, csf+18}, but a small number have been detected in the ultrafaint dwarf galaxies \citep{fn+15}.

We find that none of the stars in our sample meet the threshold to be considered CEMP stars.
However, this is not surprising given our small sample size.
For instance, \citet{pfb+14} find that CEMP stars compose 20\% of the Milky Way halo population below [Fe/H] = $-2.0$ and our sample only has two stars with [Fe/H] $< -2.0$ which makes a null CEMP detection reasonable.
Moreover, a strong CN absorption feature exists at $\sim3880$\,\AA, which is covered by the bandpass of the SkyMapper $v$ filter.
The presence of this absorption feature means that the most carbon enhanced stars are likely to be identified as more metal-rich than their true metallicity, and thus might be excluded from our target selection procedure as suggested in e.g., \citet{jkf+15}.
However, \citet{hem+18} also found no CEMP stars in their three stars with [Fe/H] $< -2.0$.
If we were to assume that the Sagittarius dSph has the same CEMP fraction as the halo, the probability of finding five stars below [Fe/H] $= -2.0$ that are not carbon-enhanced is $\sim$32\%.
While this calculation ignores sampling biases, it does suggest that a conclusive claim that CEMP stars are under-abundant in the Sagittarius dSph is difficult to argue given the small sample size.

\citet{pfb+14} find that the median corrected [C/Fe] is 0.14 for stars in the Milky Way halo with $-2.25 <\,\,\feh\,\, < -2.0$.
This value is roughly in agreement with the [C/Fe] values in our sample.
In contrast, \citet{hem+18} and \citet{hss+17} find that, over a broad metallicity range, stars in the Sagittarius dSph tend to be under-abundant in carbon relative to stars in the Milky Way halo.
This difference suggests a spread in the carbon abundances of stars at low metallicities.
Overall, results suggesting a low carbon abundance over all metallicities and a putative lack of CEMP stars suggest that early nucleosynthetic events in the Sagittarius dSph may not have been dominated by sites hypothesized to have produced CEMP stars, such as e.g., faint supernovae \citep[e.g.,][]{tun+07}.

\subsection{Searching for Metal-poor Stars using SkyMapper Photometry}
\label{sec:targetsuccess}

We observed six candidates selected as metal-poor Sagittarius dSph candidate members from SkyMapper photometry and \textit{Gaia} proper motion data (see Section~\ref{sec:targets}).
Of these six stars, four were confirmed as members by follow-up spectroscopy.
Given our success rate and the fact that only a handful of stars were found with [Fe/H] $<-2.0$ before this study \citep{bic+08, mbi+17, hem+18} in either M54 or the main body of the Sagittarius dSph, we consider our target selection procedure for metal-poor stars fairly successful.

However, it is worth inquiring into why two of our candidates were not members.
Both candidates have proper motions consistent with that of the Sagittarius dSph.
Additionally, assuming similar $T_{\text{eff}}$ values, both stars appear to have metallicities similar to Sgr-10 ([Fe/H] $\sim -1.7$) based on visually comparing the strengths of their Ca II K lines, Mg b region, and calcium triplet lines to those in the spectra of the Sagittarius dSph members.
Thus, it is likely that the non-members were simply metal-poor halo stars in the vicinity of  Sagittarius dSph.

Discriminating stars at very low metallicities ([Fe/H] $< -2.0$) from simple metal-poor stars ($-2.0 <$ [Fe/H] $< -1.0$) would likely remove some contamination from the metal-poor halo population.
Based on our spectroscopic measurements, we had mixed success at recovering only very metal-poor stars, likely because we opted to observe stars fainter then $g \sim 16$.
At these magnitudes, the public SkyMapper photometric uncertainties make a quantitative metallicity prediction unreasonable.
This fact is illustrated in Figure~\ref{fig:metallicities}.
Despite this fact, we still show a reasonable success rate in identifying metal-poor and very metal-poor stars at these magnitudes, likely because we chose to observe stars that were already predicted to have very low metallicities.
For the purposes of future work, we chose to investigate whether it would be possible to quantitatively predict metallicities for stars with sufficient precision in the public SkyMapper DR1.1 catalog.

To quantitatively relate the photometric metallicities to the overall stellar metallicities as a function of photometric precision, we generated synthetic photometry for stars with various stellar parameters.
First, we used the Turbospectrum code \citep{ap+98,p+12}, the MARCS model atmospheres \citep{gee+08}, and a line list compiled from the VALD database \citep{pkr+95, rpk+15} to generate a grid of flux-calibrated synthetic spectra.
The stellar parameters of our grid were the following: $4000 < T_{\text{eff}}\,[\text{K}] < 5700; 1 < \log\,g < 3$; $-4.0 < \text{[Fe/H]} < -0.5$. 
We then retrieved the bandpass of the SkyMapper $v$, $g$, and $i$ filters from the Filter Profile Service\footnote{http://svo2.cab.inta-csic.es/svo/theory/fps3/}, which used bandpass data from \citet{bbs+11}.
Then, we generated a library of synthetic $v$,$g$,$i$ photometry based on the methods presented in \citet{bm+12} and \citet{cv+14}.

\begin{figure*}[!htbp]
\centering
\includegraphics[width =\textwidth]{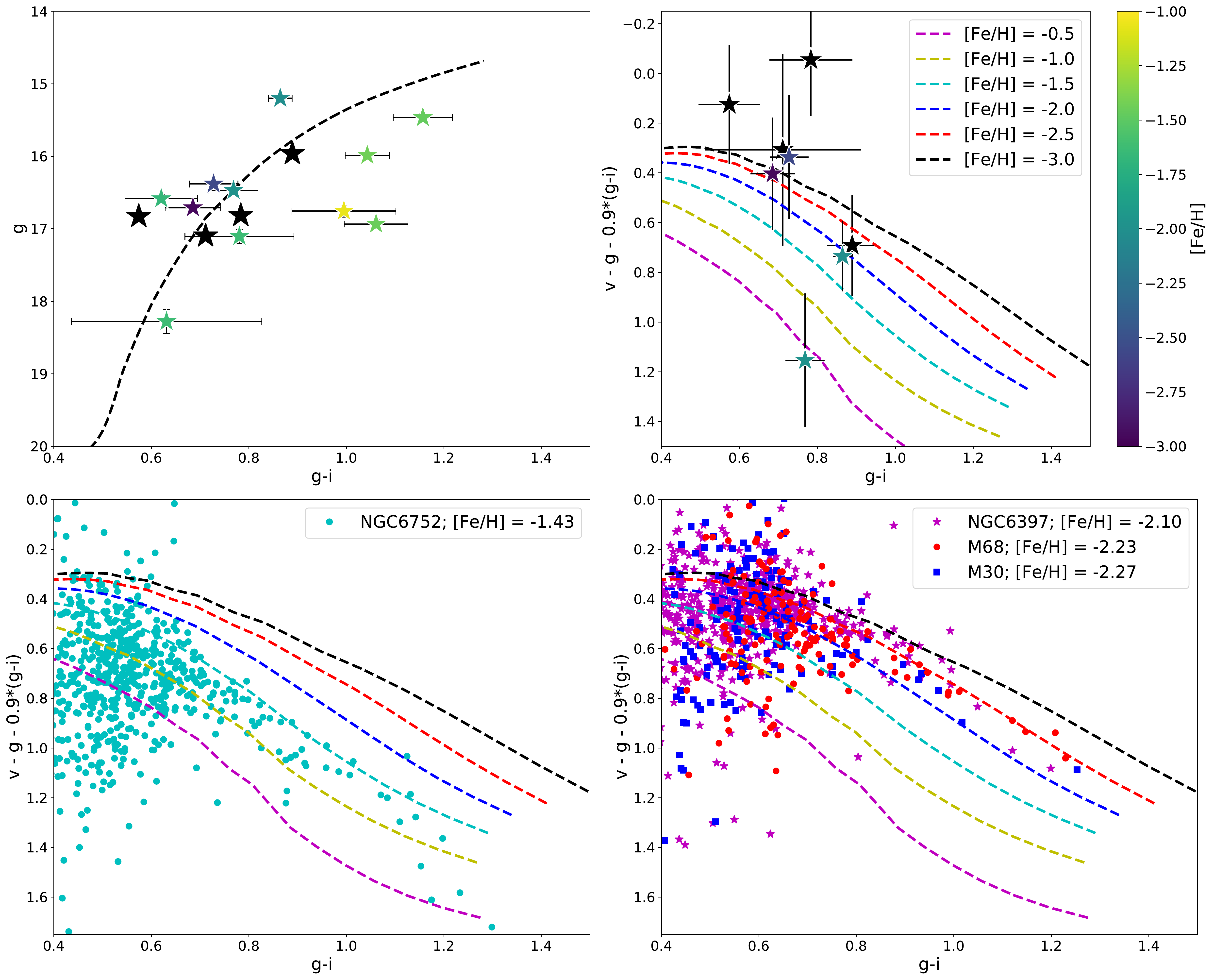}
\caption{Top left: Color-magnitude diagram of stars within 2.85$^{\circ}$ of the center of the Sgr dSph. 
Green data points are within $g-i\pm0.15$ along the overlaid Dartmouth isochrone of age 10 Gyr, [Fe/H] = $-2.0$. 
Black stars are stars with measurements from this study. 
Colored stars have measurements in \citet{hem+18}, and their metallicities are color-coded by the colorbar on the right. 
Top right: Metallicity-sensitive color-color plots using SkyMapper photometry with synthetic photometric contours and sources from the top left color-magnitude diagram. 
The stars in \citet{hem+18} with $v$ photometry available in the public SkyMapper catalog are also overlaid.
Bottom left: Metallicity-sensitive color-color plots using SkyMapper photometry with synthetic photometric contours and the metal-poor globular cluster (NGC 6752) in Figure~\ref{fig:metallicities} overlaid.
Bottom right: Metallicity-sensitive color-color plots using SkyMapper photometry with synthetic photometric contours and the three very metal-poor globular clusters (NGC 6397, M68, M30) from Figure~\ref{fig:metallicities} overlaid. }
\label{fig:synthetic}
\end{figure*}

In Figure~\ref{fig:synthetic}, we have overlaid these synthetic contours on the results from the globular cluster SkyMapper photometry.
We find that the bright stars in NGC6397, M30, and M68 lie largely between the [Fe/H] = $-2.0$ and $-2.5$ contours as expected since their metallicities are [Fe/H] = $-2.10$, $-2.23$ and $-2.27$, respectively.
As expected, fainter stars tend to fall outside these contours due to worse photometric precision.
This agreement between the synthetic contours and the globular cluster photometry suggests that future work could pre-select stars of specific metallicities for further spectroscopic study.
We note that only the most metal-poor stars in \citet{hem+18} have $v$-band SkyMapper photometry.
This is not surprising because more metal-rich stars appear fainter in the $v$ filter.
Thus, the more metal-rich stars in \citet{hem+18} are likely below the threshold for detection in the SkyMapper survey data.
When overlying our observed stars and those in \citet{hem+18} with $v$ photometry in SkyMapper, we find that all but one star is likely metal-poor, but the photometric uncertainties are sufficiently large that it is unclear which photometric metallicity contour they lie on.
Thus, this illustrates the need for precise SkyMapper photometry for target selection or solely choosing brighter stars for future candidate selection.

\section{Summary}
\label{sec:conclusion}

We present a technique for identifying metal-poor stars in dwarf galaxies using public SkyMapper photometry and {\it Gaia} proper motion data.
We obtained spectra of six stars, of which four turned out to be metal-poor members of the Sagittarius dSph.
Of the four, two have [Fe/H] $< -2.0$ and none are enhanced in carbon.
This sample builds onto the four known stars in Sgr with a metallicity below [Fe/H] $=-2.0$, showing that public proper motion and photometric data may be effectively leveraged to identify the most metal-poor stars in dwarf galaxies.
Future work will continue to implement SkyMapper photometry in tandem with {\it Gaia} proper motion data to study dwarf galaxies, both with public survey data and deep imaging of dwarf galaxies using the metallicity-sensitive SkyMapper $v$-filter (A. Chiti et al., in prep).

\acknowledgements

A.C and A.F are partially supported by NSF- CAREER grant AST-1255160 and NSF grant 1716251.
This work made use of NASA’s Astrophysics Data System Bibliographic Services, and the SIMBAD database, operated at CDS, Strasbourg, France \citep{woe+00}.

This work has made use of data from the European Space Agency (ESA) mission
{\it Gaia} (\url{https://www.cosmos.esa.int/gaia}), processed by the {\it Gaia}
Data Processing and Analysis Consortium (DPAC,
\url{https://www.cosmos.esa.int/web/gaia/dpac/consortium}). Funding for the DPAC
has been provided by national institutions, in particular the institutions
participating in the {\it Gaia} Multilateral Agreement.

The national facility capability for SkyMapper has been funded through ARC LIEF grant LE130100104 from the Australian Research Council, awarded to the University of Sydney, the Australian National University, Swinburne University of Technology, the University of Queensland, the University of Western Australia, the University of Melbourne, Curtin University of Technology, Monash University and the Australian Astronomical Observatory. SkyMapper is owned and operated by The Australian National University's Research School of Astronomy and Astrophysics. The survey data were processed and provided by the SkyMapper Team at ANU. The SkyMapper node of the All-Sky Virtual Observatory (ASVO) is hosted at the National Computational Infrastructure (NCI). Development and support the SkyMapper node of the ASVO has been funded in part by Astronomy Australia Limited (AAL) and the Australian Government through the Commonwealth's Education Investment Fund (EIF) and National Collaborative Research Infrastructure Strategy (NCRIS), particularly the National eResearch Collaboration Tools and Resources (NeCTAR) and the Australian National Data Service Projects (ANDS).

Facilities: Magellan-Baade \citep[MagE;][]{mbt+08}, SkyMapper \citep{ksb+07}

\software{Turbospectrum \citep{ap+98, p+12}, MARCS \citep{gee+08}, MOOG \citep{s+73, sks+11}, MagE CarPy \citep{k+03}, Astropy \citep{astropy}, NumPy \citep{numpy}, SciPy \citet{jop+01}, Matplotlib \citep{Hunter+07}, IRAF \citep{t+86, t+93}}

\bibliography{sgr}

\end{document}